# Cooperativity and the freezing of molecular motion at the glass transition

Th. Bauer, P. Lunkenheimer*, and A. Loidl

*Experimental Physics V, Center for Electronic Correlations and Magnetism, University of Augsburg, 86159 Augsburg, Germany*

The slowing down of molecular dynamics when approaching the glass transition generally proceeds much stronger than expected for thermally activated motions. This strange phenomenon can be formally ascribed to a temperature-dependent activation energy $E(T)$. In the present work, via measurements of the third-order nonlinear dielectric susceptibility, we deduce the increase of the number of correlated molecules $N_{corr}$ when approaching the glass transition and find a surprisingly simple correlation of $E(T)$ and $N_{corr}(T)$. This provides strong evidence that the non-canonical temperature development of glassy dynamics is caused by a temperature-dependent energy barrier arising from the cooperative motion of ever larger numbers of molecules at low temperatures.

PACS numbers: 77.22.Gm, 64.70.pm, 77.22.Ch

The glass transition is of general importance not only for the conventional silicate glasses but also for the large field of polymers, for metallic glasses, for new types of electrolytes (e.g., ionic liquids), and even for biological matter (e.g., proteins). Nevertheless, and despite many decades of intense research, this phenomenon still belongs to the greatest mysteries in condensed-matter physics and material science [1,2,3]. Since centuries, glass blowers make use of the gradual increase of viscosity when a liquid approaches the glass transition under cooling. However, in all classes of glass-forming matter the corresponding slowing down of the motion of the structural units (atoms, molecules, ions, polymer segments, etc.) proceeds significantly stronger than predicted by the Arrhenius law [1,2,3,4,5,6]. The latter is expected making the reasonable assumption that these units have to overcome an energy barrier $E$ via thermal activation to allow for translational motion and, thus, to enable viscous flow. An often proposed explanation of the found deviations from this naive expectation is an increase of the effective energy barrier $E(T)$ at low temperatures [1,2,3,4,7,8]. It may arise from an increasingly cooperative character of molecular motions when approaching the glass transition, i.e. an increase of the number of correlated molecules (or ions, polymer segments etc.) and a concomitant growth of correlation length scales [1,4,9,10]. However, until now an experimental proof of this close relation of apparent energy barrier and molecular correlation is still missing.

The experimental determination of correlation lengths in glass-forming materials is a challenging task. In principle, the measurement of a four-point correlation function, which is known to probe the cooperative length scales [11,12], would do the job. However, it seemed that the corresponding susceptibility is not directly accessible by experiment. Thus it was a breakthrough when it was shown that the nonlinear $3\omega$-harmonic of the susceptibility $\chi_3(\nu)$, which can be determined, e.g., by dielectric spectroscopy, is directly related to the four-point correlation function [13,14]. Subsequently, in a seminal work by Crauste-Thibierge et al. [15], $\chi_3$ spectra of a glass-forming liquid were determined and a significant increase of $N_{corr}$ with decreasing temperature was indeed found.

In the present work, we present $\chi_3$ of four glass-forming liquids, glycerol, propylene carbonate (PC), 3-fluoroaniline (FAN) and 2-ethyl-1-hexanol (2E1H). The departure of their $\tau(T)$ curves from Arrhenius behavior, $\tau = \tau_0 \exp[E/(k_B T)]$ ($\tau_0$: inverse attempt frequency, $E$: activation energy), differs significantly: PC [6] and FAN [16] show marked deviations and they thus can be classified as "fragile" within the commonly employed strong/fragile classification scheme of glassy matter [1,2,3,4,17]. In contrast, these deviations are much smaller for glycerol [6] and the main relaxation in 2E1H approximately follows the Arrhenius law [18]. If the growing number of correlated molecules indeed explains the increase of the apparent activation energy at low temperatures, the much stronger non-Arrhenius behavior in PC and FAN should be reflected by a stronger temperature dependence of $N_{corr}$. Moreover, relating $N_{corr}(T)$ in these glass formers to their $\tau(T)$ should shed light on the microscopic mechanism leading to the puzzling non-Arrhenius behavior of glassy matter.

The experiments were performed using a frequency-response analyzer in combination with two high-voltage boosters "HVB 300" and "HVB 4000", from Novocontrol Technologies, enabling measurements with peak voltages up to 150 or 2000 V and frequencies up to about 100 or 1 kHz, respectively. Two different ways of sample preparation were employed: i) The sample materials were put between two lapped and highly polished steel plates, forming thin films with thicknesses of 1-20 µm. No spacer materials were used, which significantly reduced the probability for electrical breakthroughs. ii) Glass-fiber spacers of 30 µm diameter were used to separate the capacitor plates. For the larger plate distances, measurements were performed using the "HVB 4000" high-voltage booster to reach sufficiently high field amplitudes. The consistency of the results from the two devices and preparation methods was carefully checked. To exclude field-induced heating effects, successive high- and low-field measurements were performed, as described in detail in Ref. 19. Field-induced sample contraction effects due to attracting forces between the capacitor plates were excluded as discussed in Ref. [19] and by comparing measurements with and without spacers.

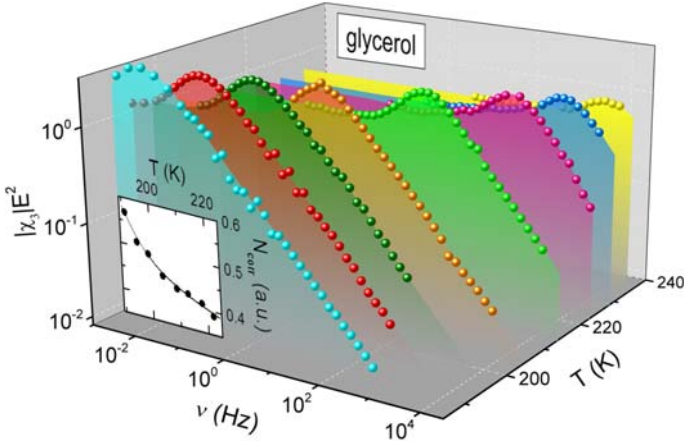

FIG. 1. Third-order harmonic component of the dielectric susceptibility of glycerol. Spectra of $|\chi_3|E^2$ are shown for various temperatures, measured at a field of 565 kV/cm. The inset shows the hump amplitude in the quantity $X$ as determined from $\chi_3$ (see text), which is proportional to the number of correlated molecules $N_{corr}$. The line is a guide to the eye.

Figure 1 shows spectra of the modulus of $\chi_3$, measured in glycerol for various temperatures when approaching the glass temperature of 185 K. (We plot $|\chi_3|$ times the squared field $E$ because of the relation $\chi_{total} = \chi_1 + \chi_3 E^2 + ...$, making this quantity dimensionless and comparable with the linear susceptibility $\chi_1$, i.e. the permittivity.) The simplest contribution to the nonlinear susceptibility arises from the well-known saturation of dipolar reorientation at high fields [20], which should lead to a plateau in $\chi_3(\nu)$ at low and a continuous decrease at high frequencies [21]. These two features are indeed found in Fig. 1, where a low-frequency plateau and a high-frequency power law, extending over more than four frequency decades, are revealed in unprecedented precision. However, superimposed to this general trend, $\chi_3(\nu)$ shows significant humps, which shift towards lower frequencies when the temperature decreases. These humps appear at frequencies somewhat lower than the well-known peaks in the linear dielectric loss spectra [6], which arise when the probing frequency matches the relaxation rate $1/(2\pi\tau)$, characterizing molecular motion. The found humps nicely agree with the results reported in the pioneering work by Crauste-Thibierge *et al.* [15]. As discussed there, within the model by Biroli and coworkers [13,14] the humped shape of the $\chi_3$ spectra is indicative of the collective nature of glassy dynamics. It was shown that the amplitude of the quantity $X = |\chi_3| k_B T / [(\Delta\varepsilon)^2 a^3 \varepsilon_0]$ is proportional to the number of correlated molecules

[13,15,22]. Here $\Delta\varepsilon$ is the relaxation strength deduced from linear spectra, $a^3$ the volume taken up by a single molecule and $\varepsilon_0$ the permittivity of free space. The inset of Fig. 1 shows the peak amplitudes of $X$, $X_{max} \propto N_{corr}$, obtained from the present experiment on glycerol. $N_{corr}(T)$ increases with decreasing temperature similar to the findings in Ref. 15.

It should be noted that recently it was shown that the humped shape of $\chi_3(\nu)$ can also be explained in the framework of other models [23,24,25]. This includes the so-called box model [23], assuming the selective heating of dynamically heterogeneous regions by the applied field, which originally was introduced for the explanation of dielectric hole-burning experiments [26]. Thus, our experimental data may also be consistent with this model. However, it was noted in Ref. 23 that the predictions of the box model and the model of refs. 13,14, concerning the link of $X$ to the number of correlated molecules, are at least formally similar.

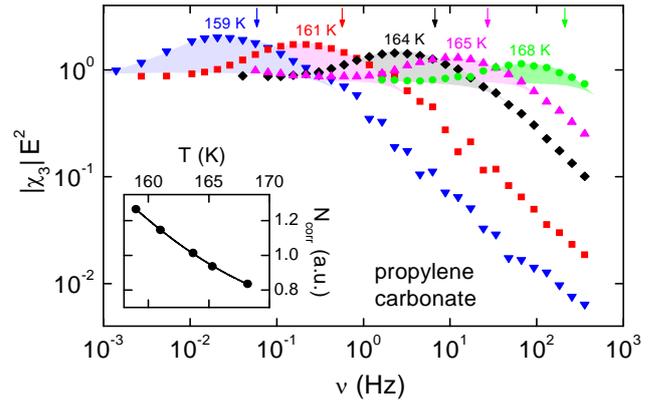

FIG. 2 (color online). Third-order harmonic component of the dielectric susceptibility of PC. Spectra of $|\chi_3|E^2$ are shown for various temperatures measured at a field of 225 kV/cm. Shaded areas show the hump. Arrows indicate the peak positions in the dielectric loss [6]. The inset shows $X_{max} \propto N_{corr}$. The line is a guide to the eye.

Figure 2 shows corresponding results for PC. Again, low-frequency saturation, high-frequency power law and the hump are observed. Remarkably, $N_{corr}(T)$ of this more fragile glass former, shown in the inset of Fig. 2, increases significantly stronger than for glycerol, despite of a much smaller temperature range. Qualitatively similar $\chi_3$ spectra were also found for FAN and its $N_{corr}(T)$ also strongly increases towards low temperatures [27]. Figure 3 presents the results for 2E1H, again revealing the characteristic spectral features including the hump. It should be noted that 2E1H belongs to the class of monohydroxy-alcohol glass formers, which are known to show unusual relaxation dynamics [18,28,29]: Their main relaxation process is not due to the structural relaxation, i.e. the motion of the molecules determining, e.g., viscous flow. Instead, it is usually ascribed to the motion of supramolecular structures (clusters) formed by several hydrogen-bonded alcohol molecules [18,29]. Interestingly, in marked contrast to the other investigated glass formers, $N_{corr}(T)$ of 2E1H, shown in the inset of Fig. 3, does not exhibit any significant temperature variation. This seems to be related to the almost Arrhenius-like $\tau(T)$ of this process [18].

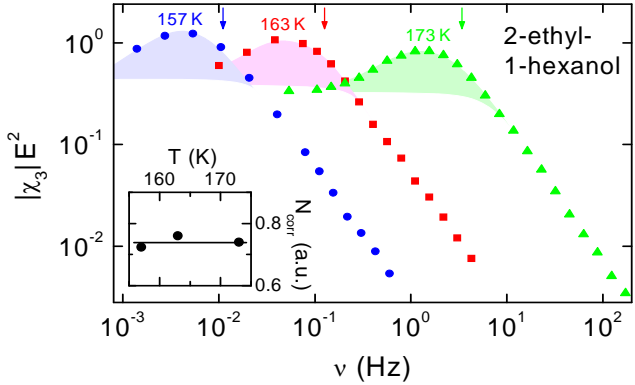

FIG. 3 (color online). Third-order harmonic component of the dielectric susceptibility of 2E1H. Spectra of $|\chi_3|E^2$ are shown for various temperatures measured at a field of 460 kV/cm. Shaded areas show the hump. Arrows indicate the peak positions in the dielectric loss [18]. The inset shows $X_{max} \propto N_{corr}$. The line demonstrates nearly constant behavior.

The inset of Fig. 4 shows the temperature dependences of the relaxation times of the investigated materials as determined by linear dielectric spectroscopy [16,18,30]. We use the commonly employed representation of $\log_{10}(\tau)$ vs. inverse temperature, which linearizes the Arrhenius law according to $\log_{10}(\tau) = \log_{10}(\tau_0) + E/(k_B T \ln 10)$. Obviously, 2E1H comes rather close to a linear increase while the other glass formers exhibit different degrees of deviation. For Arrhenius behavior, the activation energy $E$ can be determined from the slope read off in this plot. Following the notion that a temperature-dependent activation energy is responsible for the found deviations, it is suggestive to plot the derivatives of these curves, revealing information on its temperature variation. (It should be noted that, strictly speaking, the quantity $H = d(\ln \tau)/d(1/T)$ determined in this way corresponds to an apparent activation enthalpy [31].) The resulting curves are shown in the main frame of Fig. 4 (lines, right scale). Here, instead of analyzing the original data, which partly leads to excessive data scatter, we provide the derivatives of the fit curves that are shown as lines in the inset of Fig. 4. These fits were performed using the empirical Vogel-Fulcher-Tammann (VFT) function, $\tau = \tau_0 \exp[B/(T-T_{VF})]$, commonly employed to parameterize $\tau(T)$ data in glass formers [1,2,3,4]. We also tested an alternative parameterization [30] according to Mauro et al. [32], which leads to qualitatively similar results [27]. The different degrees of deviation from the Arrhenius law are closely mirrored by the temperature dependence of $d(\ln \tau)/d(1/T)$: While the activation energy is nearly constant for 2E1H, it shows strong temperature variation for PC and FAN and glycerol lies somewhere in-between these cases.

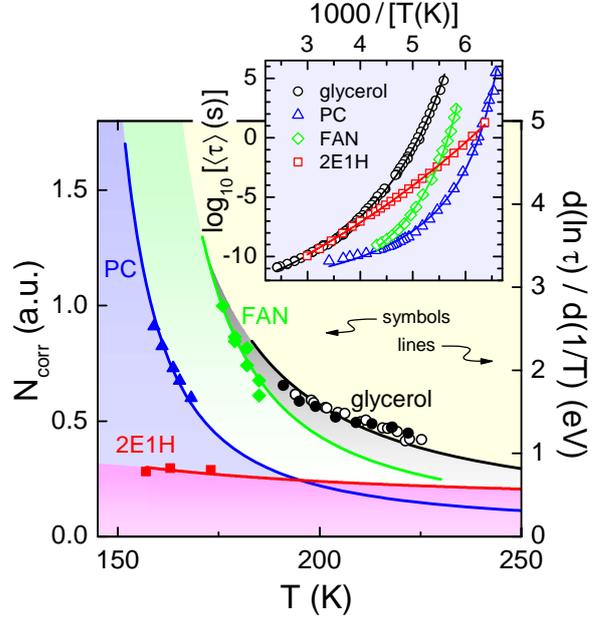

FIG. 4 (color online). Correlation of apparent activation enthalpy with number of correlated molecules. Lines: derivatives of the relaxation-time fits revealing the temperature-dependent apparent activation enthalpies (right scale). Symbols: scaled hump amplitudes of $X$, determined from $\chi_3$, which are proportional to the number of correlated molecules $N_{corr}$ (left scale). The open circles show the results on glycerol from ref. 15. The $N_{corr}$ data points (including those from ref. 15) were multiplied by separate factors for each material (glycerol: 1.15, PC: 0.72, FAN: 1.30, 2E1H: 0.39) leading to a good match with the derivative curves. (For FAN, two separate measurement runs were performed. The deviation of the results at given temperature provides an estimate of the error bars.) Inset: Temperature-dependent relaxation times of the investigated glass formers in Arrhenius representation, taken from refs. 30 (glycerol, PC), 16 (FAN), and 18 (2E1H). The lines are fits with the VFT function.

Are the markedly different temperature dependences of the activation energies revealed in Fig. 4 related to the number of correlated molecules in these glass formers? The simplest imaginable relation of the two quantities would be a linear one as assumed within the Adam-Gibbs theory of the glass transition [10]. To check for this notion, in Fig. 4 we have included the results for $N_{corr}(T)$ of the four investigated glass formers (symbols, left scale). Here we have adjusted the scaling of the ordinate to achieve a rough match of $d(\ln \tau)/d(1/T)$ and $N_{corr}$ for all materials. Additional scaling factors (glycerol: 1.15, PC: 0.72, FAN: 1.30, 2E1H: 0.39) were applied, aiming at an optimized match of $d(\ln \tau)/d(1/T)$ and $N_{corr}$ for each material (see Supplemental Material [27] for an unscaled plot). This scaling works astonishingly well: Symbols and lines agree reasonably in Fig. 4. This correlation also holds for the data on glycerol published by Crauste-Thibierge *et al.* shown as open circles in Fig. 4 [15]. This finding implies that the apparent activation energy is directly proportional to the number of correlated molecules, i.e. $H \propto N_{\text{corr}}$ (note that both ordinates start from zero, i.e. no offset was necessary to make the data match). From the definition of $N_{\text{corr}}$ it follows immediately that $H \propto N_{\text{corr}} \sim L^3$, where $L$ characterizes a dynamic length scale. Hence, the apparent effective energy barriers depend linearly on the size of the cooperatively relaxing regions, which grows on decreasing temperature. However, our experiments cannot answer the question if the length scales diverge at zero or at finite temperature. As stated earlier, the VFT law with a critical Vogel-Fulcher temperature as well as the formalism proposed by Mauro *et al.* [32], with no divergence of $\tau(T)$ at $T > 0$ K, yield similar scaling.

It should be noted here that the above considerations are primarily based on the model by Biroli and coworkers [13,14] directly relating the quantities $X$ and $N_{\text{corr}}$ but, as mentioned above, a similar relation may also be consistent with the box model [23]. However, within the latter framework the activation energy may also more directly affect the nonlinear behavior via the frequency shift of the spectral features that arises from the field-induced increase of the fictive temperatures of the heterogeneous regions [33]. A detailed analysis would be necessary to understand the exact implications of our experimental results within this model, which, however, is out of the scope of the present work.

Notably, except for 2E1H, the mentioned scaling factors applied in Fig. 4 only moderately depart from one. This indicates that the proportionality factor in $H \propto N_{\text{corr}}$ is of similar order for glycerol, PC and FAN. Thus, not only the temperature development of the activation energies for each single glass former can be explained in this way but even the absolute values of $H$ to a large extent seem to be governed by $N_{\text{corr}}$. Can the findings in 2E1H also be understood within this scenario? In comparison to the other materials, the scaling factor that is needed to make $N_{\text{corr}}$ of 2E1H match its $d(\ln\tau)/d(1/T)$ curve is significantly smaller. $N_{\text{corr}}$ is deduced from the amplitude of $X$, which itself is inversely proportional to the volume $a^3$ available to a molecule [13,15,22]. Just as for the other materials, here we have used the volume occupied by a single molecule calculated from the density and molar mass. However, as mentioned above, in monohydroxy alcohols supramolecular structures of hydrogen-bound molecules are assumed to cause the main relaxation process and these clusters can be regarded as "supermolecules", whose correlation is measured by $N_{\text{corr}}$. While the details of molecule clustering in these alcohols are a matter of debate, it is clear that simply using the single-molecule volume for $a^3$ should result in too large values of $N_{\text{corr}}$ as indeed mirrored by the found smaller scaling factor.

In conclusion, our results on the third-order harmonic susceptibility in four glass formers provide experimental evidence for an often considered but never proven conception used to explain one of the great mysteries of condensed matter: The non-canonical slowing down of molecular motion at the glass transition is caused by an increase of the number of correlated molecules when approaching the glass transition, which induces an increase of the apparent activation energy. Moreover, we find that both quantities are astonishingly simply related as the activation energy is revealed to be approximately proportional to $N_{\text{corr}}$. Finally, the similar magnitude of the proportionality factor for different molecules indicates that their activation barriers are mainly governed by the number of correlated molecules and only to a lesser extent by the type of the molecule or the interaction between them (H-bonds in glycerol and FAN *vs.* van-der-Waals interactions in PC).

We thank C. A. Angell and R. Richert for helpful discussions. This work was supported by the Deutsche Forschungsgemeinschaft via Research Unit FOR1394.

References
\* Corresponding author. Peter.Lunkenheimer@Physik.Uni-Augsburg.de
[1] P. G. Debenedetti and F. H. Stillinger, Nature **310**, 259 (2001).
[2] S. A. Kivelson and G. Tarjus, Nature Mater. **7**, 831 (2008).
[3] J. C. Dyre, Rev. Mod. Phys. **78**, 953 (2006).
[4] M. D. Ediger, C. A. Angell, and S. R. Nagel, J. Phys. Chem. **100**, 13200 (1996).
[5] G. Biroli, and J. P. Garrahan, J. Chem. Phys. **138**, 12A301 (2013).
[6] P. Lunkenheimer, U. Schneider, R. Brand, and A. Loidl, Contemp. Phys. **41**, 15 (2000).
[7] T. Hecksher, A. L. Nielsen, N. B. Olsen, and J. C. Dyre, Nature Phys. **4**, 737 (2008).
[8] J.-C. Martinez-Garcia, S. J. Rzoska, A. Drozd-Rzoska, and J. Martinez-Garcia, Nat. Commun. **4**, 1823 (2013).
[9] T. R. Kirkpatrick and P. G. Wolynes, Phys. Rev. B **36**, 8552 (1987).
[10] G. Adam, and J. H. Gibbs, J. Chem. Phys. **43**, 139 (1965).
[11] C. Dasgupta, A. V. Indrani, S. Ramaswamy, and M. K. Phani, Europhys. Lett. **15**, 307 (1991).
[12] S. Franz and G. Parisi, J. Phys.: Condens. Matter **12**, 6335 (2000).
[13] J.-P. Bouchaud and G. Biroli, Phys. Rev. B **72**, 064204 (2005).
[14] M. Tarzia, G. Biroli, A. Lefèvre, and J.-P. Bouchaud, J. Chem Phys. **132**, 054501 (2010).
[15] C. Crauste-Thibierge, C. Brun, F. Ladieu, D. L'Hôte, G. Biroli, and J.-P. Bouchaud, Phys. Rev. Lett. **104**, 165703 (2010).
[16] A. Kudlik, S. Benkhof, T. Blochowicz, C. Tschirwitz, and E. Rössler, J. Mol. Struct. **479**, 201 (1999).
[17] C. A. Angell, in *Relaxation in Complex Systems*, edited by K. L. Ngai and G. B. Wright (Office of Naval Research, Washington DC, 1985), p. 3.
[18] C. Gainaru, S. Kastner, F. Mayr, P. Lunkenheimer, S. Schildmann, H. J. Weber, W. Hiller, A. Loidl, and R. Böhmer, Phys. Rev. Lett. **107**, 118304 (2011).
[19] Th. Bauer, P. Lunkenheimer, S. Kastner, and A. Loidl, Phys. Rev. Lett. **110**, 107603 (2013).
[20] J. Herweg, Z. Phys. **3**, 36 (1920).
[21] J. L. Déjardin and Yu. P. Kalmykov, Phys. Rev. E **61**, 1211 (2000).

[22] C. Brun, F. Ladieu, D. L'Hôte, M. Tarzia, G. Biroli, and J.-P. Bouchaud, Phys. Rev. B **84,** 104204 (2011).
[23] C. Brun, C. Crauste-Thibierge, F. Ladieu, and D. L'Hôte, J. Chem. Phys. **134,** 194507 (2011).
[24] G. Diezemann, Phys. Rev. E **85**, 051502 (2012).
[25] G. Diezemann, J. Chem. Phys. **138**, 12A505 (2013).
[26] B. Schiener, R. Böhmer, A. Loidl, and R. V. Chamberlin, Science **274**, 752 (1996).
[27] See Supplemental Material for the nonlinear behavior of FAN, for an unscaled version of Fig. 4, and for an alternative analysis of the energy barriers using the fomula for $\tau(T)$, provided in Ref. 32.
[28] L.-W. Wang and R. Richert, J. Chem. Phys. **121**, 11170 (2004).
[29] C. Gainaru, R. Meier, S. Schildmann, C. Lederle, W. Hiller, E. A. Rössler, and R. Böhmer, Phys. Rev. Lett. **105**, 258303 (2010).
[30] P. Lunkenheimer, S. Kastner, M. Köhler, and A. Loidl, Phys. Rev. E **81**, 051504 (2010).
[31] J. C. Martinez-Garcia, J. Martinez-Garcia, S. J. Rzoska, and J. Hullige, J. Chem. Phys. **137**, 064501 (2012).
[32] J. C. Mauro, Y. Yue, A. J. Ellison, P. K. Gupta, and D. C. Allan, Proc. Natl. Acad. Sci. U.S.A. **106**, 19780 (2009).
[33] R. Richert, Thermochim. Acta **522**, 28 (2011).

# Cooperativity and the freezing of molecular motion at the glass transition
## Supplemental Material

Th. Bauer, P. Lunkenheimer*, and A. Loidl

*Experimental Physics V, Center for Electronic Correlations and Magnetism, University of Augsburg, 86135 Augsburg, Germany*
*e-mail: peter.lunkenheimer@physik.uni-augsburg.de*

### Third harmonic dielectric susceptibility of 3-fluoroaniline

Figure S1 shows the third-order harmonic component of the dielectric susceptibility of 3-fluoroaniline (FAN) revealing qualitatively similar behavior as for the other investigated glass formers (cf. Figs. 1-3 of main article). The inset shows the peak amplitudes of $X = |\chi_3| k_B T / [(\Delta\varepsilon)^2 a^3 \varepsilon_0]$, obtained from the present results, which are proportional to the numbers of correlated molecules $N_{corr}$ (see main article).

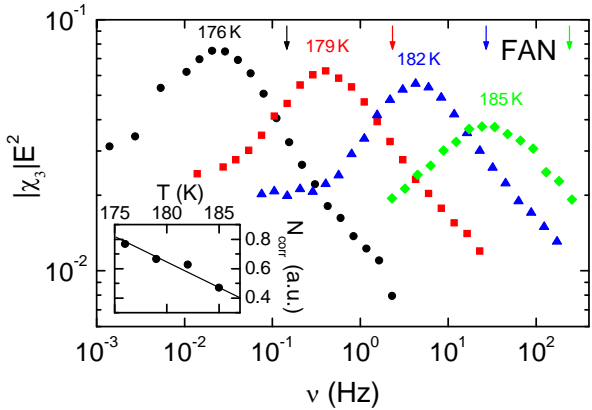

FIG. S1. Third-order harmonic component of the dielectric susceptibility of 3-fluoroaniline. Spectra of $|\chi_3|E^2$ are shown for various temperatures, measured at a field of 196 kV/cm. Arrows indicate the peak positions in the dielectric loss. The inset shows the hump amplitude in the quantity $X$ as determined from $\chi_3$ (see main article), which is proportional to the number of correlated molecules $N_{corr}$. The line is a guide to the eye.

### Comparison of $H$ and $N_{corr}$: Unscaled plot

In Fig. S2, $N_{corr}(T)$ determined from $X(\nu)$ is compared to $H(T)$ as calculated from the derivative of $\tau(T)$ in Arrhenius representation. In contrast to Fig. 4 of the main article, here $N_{corr}$ was not multiplied by a scaling factor to optimize the match of both quantities. Except for the monohydroxy alcohol 2E1H, even without separate scaling, $N_{corr}(T)$ and $H(T)$ are found to be closely related as the symbols come quite close to the corresponding lines in Fig. S2. Thus, the proportionality factor in the relation $H \propto N_{corr}$ is of similar magnitude for glycerol, PC, and FAN. Obviously, the number of correlated molecules is the dominant contribution determining the activation energy, while the molecule type and interaction is of less importance.

For the special case of the monohydroxy alcohol 2E1H, $N_{corr}(T)$ and $H(T)$ are far separated in Fig. S2 (by about a factor of 2.5), which mirrors the fact that here the dominating relaxation process arises from supramolecular structures formed by several alcohol molecules.[1]

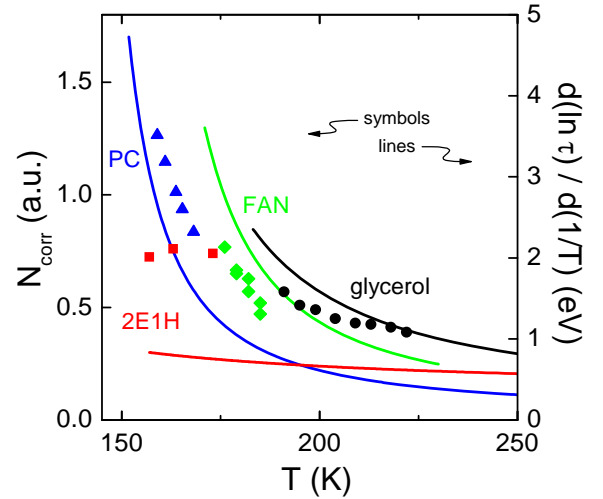

FIG. S2. Correlation of apparent activation enthalpy with number of correlated molecules. Lines: derivatives of the relaxation-time fit curves (using the VFT function) revealing the temperature-dependent apparent activation enthalpies (right scale; same data as shown in Fig. 4 of the main article). Symbols: hump amplitudes of $X$, determined from $\chi_3$, which are proportional to the number of correlated molecules $N_{corr}$ (left scale). In contrast to Fig. 4 of the main article, here no scaling factors were applied to $N_{corr}$.

## Comparison of $H$ and $N_{corr}$: Analysis with formula by Mauro *et al.*

As mentioned in the main text, the drawn conclusions are independent from the function that is used to fit the relaxation-time data. Figure S3 provides an analysis using the formula suggested by Mauro *et al.*[2] to fit the $\tau(T)$ data: $\tau = \tau_0 \exp[K/T \exp(C/T)]$. It should be noted that, in contrast to the Vogel-Fulcher-Tammann formula used in the main article, this approach does not involve any divergence of $\tau(T)$ at $T > 0$ K. Just as for the analysis with the VFT function, $N_{corr}$ is found to scale reasonably with the apparent energy barrier.

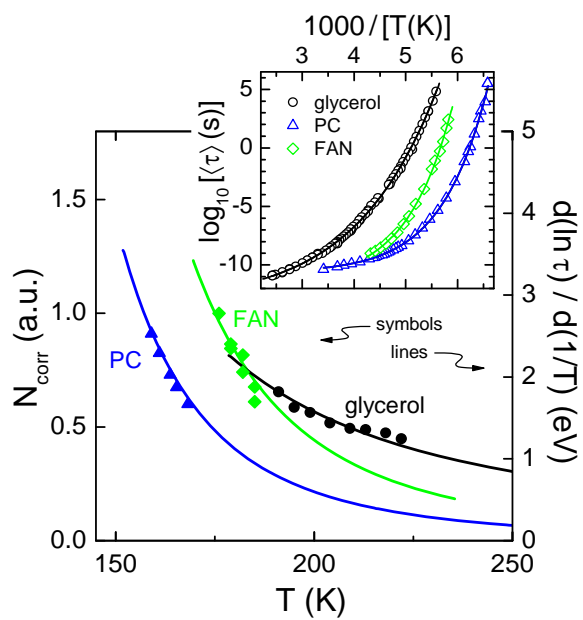

FIG. S3 Correlation of $N_{corr}$ and activation enthalpy determined using formula by Mauro *et al.* Lines: derivatives of the relaxation-time fits using the formula by Mauro *et al.*[2], revealing the temperature-dependent apparent activation enthalpies (right scale). Symbols: hump amplitudes of $X$, determined from $\chi_3$, which are proportional to the number of correlated molecules $N_{corr}$ (left scale; same data as in Fig. 4 of the main article). The data points were multiplied by the same scaling factors as in Fig. 4 of the main article. The inset shows the relaxation times, taken from refs. 3 (glycerol, PC) and 4 (FAN), in Arrhenius representation. The lines are fits with the function by Mauro *et al.*[2,3].

**References**

1. C. Gainaru, R. Meier, S. Schildmann, C. Lederle, W. Hiller, E. A. Rössler, and R. Böhmer, Phys. Rev. Lett. **105**, 258303 (2010).
2. J. C. Mauro, Y. Yue, A. J. Ellison, P. K. Gupta, and D. C. Allan, Proc. Natl. Acad. Sci. U.S.A. **106**, 19780 (2009).
3. P. Lunkenheimer, S. Kastner, M. Köhler, and A. Loidl, Phys. Rev. E **81**, 051504 (2010).
4. A. Kudlik, S. Benkhof, T. Blochowicz, C. Tschirwitz, and E. Rössler, J. Mol. Struct. **479**, 201 (1999).